\documentclass[reprint,showpacs,showkeys,aps,prd,amsfonts,eqsecnum,nofootinbib,floatfix]{revtex4-1}
\usepackage[dvips]{color}
\usepackage{multirow}
\usepackage{hyperref}
\usepackage{epsfig}
\usepackage{rotate}
\usepackage{rotating}
\usepackage{multirow}
\usepackage{longtable}
\usepackage[T1]{fontenc}
\usepackage{array}
\usepackage{graphicx,amssymb,amsfonts,amsmath}
\usepackage{feynmp-auto}

\def\gsim{\lower0.5ex\hbox{$\:\buildrel >\over\sim\:$}}
\def\lsim{\lower0.5ex\hbox{$\:\buildrel <\over\sim\:$}}
\def\fo{\hbox{{1}\kern-.
25em\hbox{l}}}

\usepackage{colortbl}
\newcommand{\bea}{\begin{eqnarray}}
\newcommand{\eea}{\end{eqnarray}}

\newcommand{\be}{\begin{equation}}
\newcommand{\ee}{\end{equation}}

\def\bsp#1\esp{\begin{split}#1\end{split}}
\def\bpm{\begin{pmatrix}}
\def\epm{\end{pmatrix}}

\def\beq{\begin{equation}}
\def\eeq{\end{equation}}
\def\eq{\end{equation}}
\def\to{\rightarrow}

\def\bsg{\ifmmode B\to X_s\gamma\else $B\to X_s\gamma$\fi}
\def\bsll{\ifmmode B\to X_s\ell^+\ell^-\else $B\to X_s\ell^+\ell^-$\fi}
\def\bstt{\ifmmode B\to X_s\tau^+\tau^-\else $B\to X_s\tau^+\tau^-$\fi}
\def\shat{\ifmmode \hat{s}\else $\hat{s}$\fi}

\def\s2b{s_{2\beta}}

\def\u1{G_{SM}\otimes U(1)^{\prime}}

\begin{document}

\begin{flushright}  \mbox{\normalsize \rm HIP-2017-15/TH}  \end{flushright}

\title{Froggatt-Nielsen mechanism in a model with $SU(3)_c\times SU(3)_L \times U(1)_X$ gauge group}

\author{Katri Huitu\footnote{Email: \href{katri.huitu@helsinki.fi}{katri.huitu@helsinki.fi}}}
\author {Niko Koivunen\footnote{Email: \href{niko.koivuneni@helsinki.fi}{niko.koivunen@helsinki.fi}}}

\affiliation{Helsinki Institute of Physics and Department of Physics, 
P.O. Box 64 (Gustaf H\"allstr\"omin
  katu 2), FI-00014 University of Helsinki, Finland,}

\date{\today}

\begin{abstract}
The models with the gauge group $SU(3)_c\times SU(3)_L \times U(1)_X$ (331-models) have been advocated to 
explain why there are three fermion generations in Nature. 
As such they can provide partial understanding of the flavour sector.
The hierarchy of Yukawa-couplings in the Standard Model is another puzzle which remains without
compelling explanation.
We propose to use Froggatt-Nielsen -mechanism in a 331-model to explain both fundamental problems.
It turns out that no additional representations in the scalar sector are needed to take care of this. 
The traditional 331-models predict unsuppressed scalar flavour changing neutral currents at tree-level. We show that they are strongly suppressed in our model.


\end{abstract}

\maketitle

\renewcommand{\theequation}{\thesection\arabic{equation}}
\noindent 
{\it Introduction. }
Understanding the number of generations has been attempted using models with an extended gauge group $SU(3)_c\times SU(3)_L \times U(1)_X$ \cite{Singer:1980sw,Valle:1983dk,Foot:1994ym, Pisano:1991ee, Frampton:1992wt}. 
In these so-called 331-models the chiral anomalies are cancelled if the numbers of triplets and antitriplets are equal.
This is possible only if the number of generations is three.
Although the 331-models give a possible explanation for the number of generations, they do not shed any light on vastly different masses among the generations. 
One possibility is to explain the masses dynamically within the Froggatt-Nielsen -mechanism (FN) \cite{Froggatt:1978nt}. 

So far no attempt to combine these two approaches has been made to the best of our knowledge.
We propose here a model, where a combination of scalars in the model emulate the flavon of the FN mechanism\footnote{
In frameworks with Higgs doublets  two electrically neutral fields have been used for similar purpose, see \cite{Babu:1999me}.}.
We emphasize that this requires special properties from the model, as explained later.

We construct here a model, where the gauge symmetry of a 331-model is broken by three scalar triplets, which is the minimum number of triplets to break the symmetry and generate tree-level masses for all the charged fermions.
An attractive feature of our model is that the $SU(3)_L\times U(1)_X$ breaking scale is at the same time the scale of flavour symmetry breaking.
\\

\noindent
{\it Particle content in the 331-model.}
The symmetry breaking of a 331-model is more involved than in the case of the Standard Model (SM), since two stage breaking is needed: $SU(3)_L \times U(1)_X \rightarrow SU(2)_L \times U(1)_Y \rightarrow U(1)_{em}$.

Using the three diagonal generators $T_3,\,T_8$ and $X$, the electric charge of particles can be defined as
\begin{equation}
Q=T_3+\beta T_8+X.
\end{equation}
Different values of $\beta$ have been considered in literature. 
$\beta=\pm 1/\sqrt{3}$ \cite{Singer:1980sw,Valle:1983dk,Foot:1994ym} and $\beta=\pm \sqrt{3}$ \cite{Pisano:1991ee,Frampton:1992wt} lead to integer charges for gauge bosons and  scalars, see {\it e.g.} \cite{Cao:2016uur}.
The proposed mechanism will be found for 
$\beta = \pm\frac{1}{\sqrt{3}}$, with which there are no exotic electric charges. 

With $\beta = -\frac{1}{\sqrt{3}}$ only two types of scalar triplets with neutral components can be formed.
In order to generate tree-level masses for all the fermions, three scalar triplets are needed \cite{Valle:1983dk}.
In our model we use
\begin{eqnarray}
&&\eta=\left(\begin{array}{c}
\eta^{+}\\
\eta^{0}\\
{\eta'}^{+}
\end{array}
\right), \quad
\rho=\left(\begin{array}{c}
\rho^{0}\\
\rho^{-}\\
{\rho'}^{0}
\end{array}
\right), \quad
\chi=\left(\begin{array}{c}
\chi^{0}\\
\chi^{-}\\
{\chi'}^{0}
\end{array}
\right),\label{scalar triplets}
\end{eqnarray}
where $\eta\sim (1,3,\frac{2}{3})$ and $\rho,\chi\sim (1,3,-\frac{1}{3})$.  The numbers in the parantheses label the transformation properties under the gauge group  $SU(3)_{c}\times SU(3)_{L}\times U(1)_{X}$. 

All  the neutral fields can in general develop a non-zero vacuum expectation value (VEV).  
Degenerate minima are related to each other  by $SU(3)_L$ rotation. Thus, the most general combination of VEVs can be written as
\begin{equation}
\langle\eta^0\rangle=\frac{v'}{\sqrt{2}},\,\langle\rho^0\rangle=\frac{v_1}{\sqrt{2}},
\, \langle\rho'^0\rangle=\frac{v_2}{\sqrt{2}},\, \langle\chi'^0\rangle=\frac{u}{\sqrt{2}}.
\label{vacuum}\end{equation}
Here $v_2$ and $u$ break the $SU(3)_L\times U(1)_X \to SU(2)_L\times U(1)_Y$, and thus we expect $v_2,u \gg v',v_1$.

As required by anomaly cancellation, the same number of triplets and antitriplets is needed. 
Here this is achieved 
by assigning the leptons and one of the quark families to $SU(3)_{L}$  -triplets, while two quark families belong to antitriplets. 
The leptons are given by
\begin{eqnarray}
&&L_{L,i}=\left(\begin{array}{c}
\nu_{i}\\
e_{i}\\
\nu'_{i}
\end{array}
\right)_{L}\sim (1,3,-\frac{1}{3}),\quad  e_{R,i}\sim (1,1,-1),
\label{331 leptons}
\end{eqnarray}
where $i=1,2,3$. The fields $\nu'_i$  are new neutral leptons. 
For $\beta=\pm\sqrt{3}$, in the minimal case the triplet includes charged lepton $e^c_{i}$ instead of $\nu_i'$, and no charged singlets are required. However, as discussed later, the scalar structure of such model does not suit our purposes.

We choose to assign first quark generation into triplet and the second and the third  into antitriplet:
\begin{eqnarray}
&&Q_{L,1}=\left(\begin{array}{c}
u_{1}\\
d_{1}\\
U
\end{array}
\right)_{L}\sim (3,3,\frac{1}{3}),\\
&&Q_{L,2}=\left(\begin{array}{c}
d_{2}\\
-u_{2}\\
D_{1}
\end{array}
\right)_{L},
Q_{L,3}=\left(\begin{array}{c}
d_{3}\\
-u_{3}\\
D_{2}
\end{array}
\right)_{L}\sim (3,3^{\ast},0),\nonumber\\
&&  u_{R,i},  U_{R}\sim (3,1,\frac{2}{3}),\,  d_{R,i}, D_{R,1}, D_{R,2}\sim (3,1,-\frac{1}{3}),\nonumber
\end{eqnarray}
 where $i=1,2,3$.
We have introduced new quarks $D_{1}$ and  $D_{2}$ with electric charge $-1/3$ and $U$ with electric  charge $2/3$. 
\\

\noindent
{\it Froggatt-Nielsen -mechanism.}
Froggatt and Nielsen proposed a new symmetry, so-called flavour symmetry ({\it e.g.} U(1) or $Z_N$), to be responsible for the huge differences in fermion masses in the SM \cite{Froggatt:1978nt}.
The FN  charge assignment is  such that the SM Yukawa-couplings are forbidden, but the following effective operator is allowed:
\begin{equation}
c^{f}_{ij}\left(\phi/\Lambda\right)^{n^f_{ij}}\bar{\psi}_{L,i}^f H f_{R,j}+h.c.,
\end{equation}
where $c^f_{ij}$ is a dimensionless order-one number, $\Lambda$ is the scale of new physics, $\psi_i$, $f_i$ are the SM fermions, $H$ is the SM Higgs doublet and $\phi$ is a complex scalar field called flavon.  
The SM Yukawa couplings 
are generated as effective couplings when flavon gets a VEV
\begin{equation}
y^f_{ij}=c^f_{ij}\left(\langle\phi\rangle/\Lambda\right)^{n^f_{ij}}.
\end{equation}

Our purpose here is to mimick the gauge singlet flavon by a combination of the existing scalars in the model.
\\

\noindent
{\it Froggatt-Nielsen -mechanism in the 331-model.}
We introduce a global  $U(1)_{FN}$-symmetry to our model. Fermions and scalars are charged under it.
Two of the three triplets in Eq.\eqref{scalar triplets} carry necessarily the same $U(1)_X$ charge.
For our purposes this is crucial, since  the combination $\rho^{\dagger}\chi$ is a gauge singlet. 
Therefore, it can act as a flavon field when developing a VEV, if the FN charge of the combination does not vanish.
We take the FN charge of scalar triplets as follows:
\be
q_{\eta}=-1,\quad q_\rho =1,\quad q_\chi = 0.
\label{FN charges of scalars}
\ee
Note that the scalar representations generating tree-level masses for fermions in models with $\beta=\pm\sqrt{3}$ \cite{Pisano:1991ee,Frampton:1992wt}  cannot form an $SU(3)_L\times U(1)_X$ singlet with an FN charge.
Thus in these models, an enlarged scalar sector is needed for FN mechanism. 

The effective flavon obtains a  non-zero  vacuum expectation value  
\be
\langle\rho^{\dagger}\chi\rangle=(v_2 u)/2. 
\ee
The relevant effective operator  is now 
\begin{equation}\label{331 effective operator}
(c_s^{f})_{ij}\left(\frac{\rho^{\dagger}\chi}{\Lambda^2}\right)^{(n^s_f)_{ij}}\bar{\psi}_{L,i}^f S f_{R,j}+h.c.,
\end{equation}
where $(c_s^{f})_{ij}$  is a dimensionless order-one number. 
The fermion (anti)triplets and singlets $\bar{\psi}_{L,i}^f$  and $f_{R,j}$ have FN charges  $q(\bar{\psi}_{L,i}^f)$ and  $q(f_{R,j})$, respectively.
$S$ denotes any of the three scalar triplets $\eta$, $\rho$ or $\chi$ with which the combination is a gauge singlet.
The $(n^s_f)_{ij}$ is determined by the FN charge assignment, 
\begin{equation}
(n^s_f)_{ij}=\left[q(\bar{\psi}_{L,i}^f)+q(f_{R,j})+q(S)\right].
\end{equation}
Here $v_2\sim u$ is responsible for both the $SU(3)_L\times U(1)_X$ breaking and of the flavour symmetry breaking, while $\Lambda $ is the scale of new physics.
For simplicity we have chosen to work with charge assignments  that keep $(n^s_f)_{ij}$ non-negative. If $(n^s_f)_{ij}$ were negative, we would include  $\chi^{\dagger}\rho$ instead of $\rho^{\dagger}\chi$.

As one expands (\ref{331 effective operator}) around the minimum one obtains
\begin{eqnarray}
&&(y^f_s)_{ij}\bar{\psi}_{L,i}^f(S+\langle S\rangle) f_{R,j}
+(n^s_f)_{ij}(y^f_s)_{ij} \label{331 FN}\\
&&\times\left[\frac{{\rho'}^{0\ast}}{v_2}+\frac{{\chi'}^0}{u}
+\frac{v_1 {\chi}^0 }{v_2 u}\right]\sqrt{2}\bar{\psi}_{L,i}^f\langle S\rangle f_{R,j}+h.c.+\cdots,
\nonumber
\end{eqnarray} 
where we have kept only the renormalizable contributions. 
The first term gives the fermion masses and the  usual Yukawa interaction: 
\begin{equation}
(y^f_s)_{ij}=(c^f_s)_{ij}\left(\frac{v_2 u}{2\Lambda^2}\right)^{(n^s_f)_{ij}} \equiv (c^f_s)_{ij} \epsilon ^{(n^s_f)_{ij}}.
\end{equation}
The other term of \eqref{331 FN} is the artefact of the Froggatt-Nielsen -mechanism introducing a source of flavour violation into the model. 
The flavour violating term is  
suppressed by the $SU(3)_L$ breaking scale, $u$ and $v_2$,  in the case of the SM fermions. 
Therefore, the flavour violation coming from this additional term is negligible.  
\\

\noindent
{\it Symmetry breaking and the bosonic sector.}
The most general scalar potential of our model, consistent with Froggatt-Nielsen -mechanism is:
\begin{eqnarray}
&& V_{\textrm{FN}}=\mu^2_1 \eta^{\dagger}\eta+\mu_2^2 \rho^{\dagger}\rho+\mu_3^{2}\chi^{\dagger}\chi
+\lambda_1 (\eta^{\dagger}\eta)^2+\lambda_2 (\rho^{\dagger}\rho)^2\nonumber\\
&&+\lambda_{3}(\chi^{\dagger}\chi)^2 +\lambda_{12} (\eta^{\dagger}\eta)(\rho^{\dagger}\rho)+\lambda_{13} (\eta^{\dagger}\eta)(\chi^{\dagger}\chi)\nonumber\\
&&+\lambda_{23}(\rho^{\dagger}\rho)(\chi^{\dagger}\chi)+\widetilde{\lambda}_{12} (\eta^{\dagger}\rho)(\rho^{\dagger}\eta)+\widetilde{\lambda}_{13} (\eta^{\dagger}\chi)(\chi^{\dagger}\eta)\nonumber\\
&& +\widetilde{\lambda}_{23}(\rho^{\dagger}\chi)(\chi^{\dagger}\rho)+\sqrt{2}f(\epsilon_{ijk}\eta^{i}\rho^{j}\chi^{k}+h.c.).\nonumber
\end{eqnarray}

One scalar and seven pseudoscalar massless degrees of freedom are needed
for the missing polarization states of the gauge bosons of the model, namely
the $Z,W^\pm$ of the SM, new heavy charged gauge bosons, a
heavy neutral gauge boson, 
and a non-Hermitian heavy neutral gauge boson.
Thus, in the physical spectrum one has four scalars, two pseudoscalars, and two charged Higgs bosons.

The global $U(1)_{FN}$ -symmetry is spontaneusly broken by the scalar field VEVs, and one of the pseudoscalars will be a massless Goldstone boson.   
In order to give a mass for the Goldstone boson, we add the following soft FN symmetry breaking term to the potential:
\begin{equation}
V_{\textrm{soft}}=b(\rho^{\dagger}\chi)+h.c.
 \end{equation}
One of the scalar eigenvalues is given by light VEVs corresponding to the experimentally detected Higgs boson mass.
The remaining three scalars  as well as the charged Higgses are heavy with masses proportional to large VEVs and decouple.
One of the physical pseudoscalars is heavy, and the mass of the pseudo-Goldstone boson, $A_2$, is proportional to the soft parameter $b$.
\\

\noindent
{\it Fermion masses in the model.}
The charged lepton masses are generated by the term:
\begin{equation}
\mathcal{L}\supset y^e_{ij}\bar{L}'_{L,i}\eta e'_{R,j}+h.c.,
\end{equation}
where prime denotes gauge eigenstate. The mass matrix is given by
$$
\mathcal{L}_{\ell -mass}
=m^e_{ij}\bar{e}'_{L,i}e'_{R,j}
+h.c.\,\, {\textrm{with}}\,\,
m^e_{ij}=(y^{e}_\eta)_{ij}\frac{v'}{\sqrt{2}}.$$
The mass matrix is proportional to the Yukawa-coupling matrix.
Both will be diagonalized simultaneously.
As for neutrinos, with the fields in \eqref{331 leptons}, one of the neutrinos gets mass only radiatively, at the same time raising degeneracy from the masses of the other two.
Such an approach is demonstrated {\it e.g.} in \cite{Valle:1983dk}.
This is the approach we adopt here. Tree-level masses can also be produced, if new neutral fermions $N_{R,i}$  are introduced \cite{Huitu:2019mdr}.

The up-type quark masses are generated by the following terms:
\begin{eqnarray}
\label{u-L}
\mathcal{L}\supset &&\sum_{\gamma=1}^{4}(y^u_\rho)_{1\gamma}\bar{Q}'_{L,1}\rho ~u'_{R,\gamma}+\sum_{\gamma=1}^{4}(y^u_\chi)_{1\gamma}\bar{Q}'_{L,1}\chi ~u'_{R,\gamma}\nonumber\\
&&+\sum_{\alpha=2}^{3}\sum_{\gamma=1}^{4}(y^u_{\eta^{\ast}})_{\alpha\gamma}\bar{Q}'_{L,\alpha}\eta^{\ast} ~u'_{R,\gamma}+h.c.
\end{eqnarray}
The up-quark mass matrix elements in the gauge eigenbasis  are given by
\begin{eqnarray}
&&\mathcal{L}_{up-mass}=\bar{u}'_{L} m^u u'_R+h.c.\quad\textrm{with}\\
&&m^u_{1\gamma}=\frac{v_1}{\sqrt{2}} (y^{u}_{\rho})_{1\gamma}, \quad m^u_{\alpha\gamma}=-\frac{v'}{\sqrt{2}} (y^{u}_{\eta^{\ast}})_{\alpha\gamma},\\
&&m^u_{4\gamma}=\frac{v_2}{\sqrt{2}} (y^{u}_{\rho})_{1\gamma}+\frac{u}{\sqrt{2}} (y^{u}_{\chi})_{1\gamma},
\end{eqnarray}
where $\alpha=2,3$ and $\gamma=1,2,3,4$. The  down-quark mass matrix and the terms that generate it  are given in the Appendix.

In 331-setting the hierachy of the matrix elements is determined by the FN charge assignment and different VEVs, in contrast to traditional FN mechanism, where the hierarchy is set solely by the charge assignment. 
The hierarchy of VEVs, $u, v_2\gg v_1,v'$, greatly affects the hierarchy in the matrix. 
 We rewrite the elements of the mass matrix to clarify the hierarchy:
\begin{eqnarray}
m^u_{1\gamma}&=&\frac{v'}{\sqrt{2}} \left[\frac{v_1}{v'}(c_{\rho}^{u})_{1\gamma}\right]\epsilon^{a_1^{u}+q(u_{R,\gamma})},\\
m^u_{\alpha\gamma}&=&\frac{v'}{\sqrt{2}} \left[-(c_{\eta^{\ast}}^{u})_{\alpha\gamma}\right]\epsilon^{a_\alpha^{u}+q(u_{R,\gamma})},\quad \alpha=2,3,\\
m^u_{4\gamma}&=&\frac{v'}{\sqrt{2}} \left[(c_{\rho}^{u})_{1\gamma}\epsilon^{q(\rho)-q(\chi)}
+(c_{\chi}^{u})_{1\gamma}\frac{u}{v_2}\right]\epsilon^{a_4^{u}+q(u_{R,\gamma})},\nonumber
\end{eqnarray}
where the  quantities in square brackets are order-one numbers, and therefore the hierarchy is completely set by the powers of $\epsilon$. The  $a^{u}_{\gamma}$ are :
\begin{eqnarray}
&&a^{u}_1= q(\bar{Q}_{L,1})+q(\rho),\quad a^{u}_\alpha= q(\bar{Q}_{L,\alpha})+q(\eta^{\ast}),\quad \alpha=2,3,\nonumber\\
&&a^{u}_4= (\log \epsilon)^{-1}\log\left(\frac{v_2}{v'}\right)+q(\bar{Q}_{L,1})+q(\chi).\label{effective charges up}
\end{eqnarray}
The difference between two symmetry breaking scales manifests itself as effective left-handed charges,
analogous to FN  charges of the left-handed fermion doublets in the original FN mechanism.  
This way, the textures of the diagonalization matrices can be  easily obtained. 
\\

\noindent
{\it Suppression of flavour changing neutral currents.}
As explicitly shown in Eq.\eqref{u-L}, the 
quarks couple to multiple scalar triplets, and 331-models generally predict scalar mediated 
flavour changing neutral currents
(FCNC) of quarks at tree-level  \cite{Glashow:1976nt}, without a natural suppression mechanism.
Although this is not the main motivation in our current work, we shortly discuss the suppression of FCNCs in the model, and return to this in another work \cite{Huitu:2019kbm}.
The gauge sector in our model remains the same as in other 331-models, and 
the GIM-mechanism  \cite{Glashow:1970gm} 
has been studied in 331-models previously \cite{Georgi:1978bv}.

Although most of the scalars are heavy, the FCNC mediated by the lightest Higgs, $h$, or by the pseudo-Goldstone boson $A_2$ can be large.
Here we assume that $b$ is large, which suppresses the FCNC by $A_2$, and concentrate on the Higgs boson. 

In our model the charged lepton generations are assigned to same representations.
The only flavour violating leptonic couplings to the neutral  scalars are due to the FN -mechanism, coming from the second term in Eq.\eqref{331 FN}. However,  it is suppressed by the large $SU(3)_L$-breaking VEVs, and is negligible.

The Yukawa interactions of quarks  with the lightest Higgs boson can be written as
\begin{eqnarray}
\mathcal{L}_{q,Yukawa}
&=& \frac{1}{\sqrt{2}}\bar{u}_{L}U_L^u{\Gamma'}^u_{h}U_{R}^{u\dagger}u_{R}h+\frac{1}{\sqrt{2}}\bar{d}_{L}U_{L}^d{\Gamma'}^d_{h}U_{R}^{d\dagger}d_{R}h,\nonumber
\end{eqnarray}
where the quark gauge eigenstates have been rotated to the physical states by $U_{L,R}^{u,d}$ matrices.

We ignore here the heavily suppressed FN contributions to Yukawa interactions of the  quarks.
The physical Yukawa couplings are $\Gamma^u_h=U_L^u{\Gamma'}^u_{h}U_{R}^{u\dagger}$ and  $\Gamma^d_h=U_L^d{\Gamma'}^d_{h}U_{R}^{d\dagger}$.

By  using  the explicit form for the coupling matrices and approximating the $h$ eigenvector, we can write the coupling matrix in the form:
\begin{eqnarray}
(\Gamma^u_h)_{ij} & = & \sqrt{2}\frac{m_j}{v_{SM}}\left[\delta_{ij} 
+\alpha_1 (U_L^u)_{i1}(U_L^{u\dagger})_{1j}
-(U_L^u)_{i4}(U_L^{u\dagger})_{4j}\right. \label{fcnc up2}\nonumber\\
&&+\left. \alpha_2 (U_L^u)_{i1} (U_L^{u\dagger})_{4j}
+\alpha_3 (U_L^u)_{i4} (U_L^{u\dagger})_{1j}\right],
\end{eqnarray}
where $\alpha_i=\mathcal{O}(v_{light}/v_{heavy}),\,i=1,2,3$, with  $v_{\textrm{light}}=v',v_1$,  and $v_{\textrm{heavy}}=u, v_2$.
The corresponding couplings $(\Gamma^d_h)_{ij}$ follow a similar pattern. We assume
\begin{equation}\label{mass matrix hierarchy}
 m^q_{i,j}\leq m^q_{i+1,j},
\end{equation}
which can be ensured by a proper choice of FN charges. Rotation matrix elements then satisfy $(U^{u}_L)_{ij}\sim \epsilon^{\lvert a^{u}_i-a^{u}_j}\rvert$, which provides additional suppression to FCNCs.   

As a concrete example we set 
$\epsilon=0.23$, and consider the following FN charge assignments for the left-handed 
quark triplets:
 $q(Q^c_{L,1})=2$, $q(Q^c_{L,2})=1$, $q(Q^c_{L,3})=-1$. This charge assignment will produce the correct texture for the CKM-matrix.
The FN charges of the right-handed quarks are: $q(u_{R,1})=5$,  $q(u_{R,2})=2$,  $q(u_{R,3})=0$, $q(u_{R,4})=0$,  $q(d_{R,1})=7$,  $q(d_{R,2})=5$, $q(d_{R,3})=4$, $q(d_{R,4})=3$ and $q(d_{R,5})=2$.
The up-type Yukawa-matrix texture becomes:
\begin{equation}\label{Up-type Yukawa matrix texture}
\Gamma^u_h\sim\left(\begin{array}{llll}
y_u & y_c[\alpha \epsilon^1] & y_t[\alpha \epsilon^x] & \epsilon^2\\
y_u[\alpha \epsilon^1] & y_c & y_t[\alpha \epsilon^x] & \epsilon^2\\
y_u[\alpha \epsilon^x] & y_c[\alpha \epsilon^x] & y_t & 1\\
y_u[\alpha ] & y_c [\alpha ] & y_t[ \epsilon^x] & \epsilon^x
\end{array}\right),
\end{equation}
where   $x=  (\log \epsilon)^{-1}\log\left(\frac{v'}{v_2}\right)-2 \geq 0$ due to Eq.(\ref{mass matrix hierarchy}), and $\alpha=\mathcal{O}(v_{light}/v_{heavy})$. The diagonal couplings of the SM quarks are the SM Yukawa-couplings and  the off-diagonal elements are suppressed by  the ratio of the  two VEV scales and the powers of $\epsilon$. Similar texture applies for the down-type Yukawa-matrix. We find that the heavy VEVs$\sim {\cal{O}}$(5 TeV) can sufficiently suppress all the meson mixing constraints $M^0-\bar{M}^0$, $M=K,B_d,B_s,D$, which provide the tightest constraints in our case.

The texture in Eq.\eqref{Up-type Yukawa matrix texture} produces exotic quark masses that are suppressed compared to the $SU(3)_L$-breaking scale: $m_U\sim \epsilon^2 v_{heavy}$,  $m_{D_1}\sim \epsilon^3 v_{heavy}$ and  $m_{D_2}\sim \epsilon^0 v_{heavy}$.  
The experimental mass limit for exotic quarks is ${\mathcal O}$(1 TeV) \cite{Patrignani:2016xqp}, and
thus $v_{heavy}$ should be somewhat larger than required by the suppression of FCNCs. The texture for the SM quark masses are: $m_u\sim \epsilon^8 v_{light}$, $m_c\sim \epsilon^4 v_{light}$, $m_t\sim \epsilon^0 v_{light}$, $m_{d}\sim \epsilon^8 v_{light}$, $m_s\sim \epsilon^5 v_{light}$ and $m_b\sim \epsilon^2 v_{light}$.

The CKM-matrix is not a square matrix in this model, but a $4\times 5$-matrix.
 \begin{equation}
\mathcal{L}_{gCC}=\frac{g_{3}}{\sqrt{2}}\bar{u}_L\gamma^{\mu}V_{CKM}^{331}d_L {W}^{+}_{\mu}+h.c.
\end{equation}
The  CKM-matrix texture in our example is:
\begin{equation}
V_{CKM}^{331}\sim \left(\begin{array}{ccccc}
1            & \epsilon^{1} & \epsilon^{3}        & \epsilon^{1}\alpha    & \epsilon^{3}\alpha\\
\epsilon^{1} & 1            & \epsilon^{2}        & \alpha                & \epsilon^{2}\alpha\\
\epsilon^{3} & \epsilon^{2} & 1                   & \epsilon^{-2}\alpha   & \alpha\\
\alpha       & \alpha       & \epsilon^{-1}\alpha & \epsilon^{-4}\alpha^2 & \epsilon^{-2}\alpha^2\\
\end{array}\right).
\end{equation}
The $3\times 3$-block in the upper-left corner corresponds to the CKM-matrix of the Standard Model. The $W_\mu$-boson couples to the exotic quarks and they contribute to the neutral meson mixing. We find that the $W$-mediated meson mixing is always subleading to Higgs mediated.

 As a numerical example we set the $SU(3)_L$-breaking VEVs  to be $u=48$ TeV and $v_2=55$ TeV, and the $SU(2)_L$-breaking VEVs are $v_1=100$ GeV and $v'=237.05$ GeV. The exact quark mass matrices are given in the appendix and they produce the experimentally measured quark masses \cite{Patrignani:2016xqp}. The exotic quark masses become: $m_U=5$ TeV, $m_{D_1}=1.295$ TeV  and $m_{D_2}=50.9$ TeV. The physical up-type Yukawa-coupling matrix is:
 \begin{equation}
 \resizebox{0.47 \textwidth}{!} 
{
$ 
\Gamma^u_h=\left(\begin{array}{cccc}
1.7\times 10^{-5} & 2.9\times 10^{-9} & -5.0\times 10^{-5} & -2.5\times 10^{-2}\\
-7.7\times 10^{-11} & 7.3\times 10^{-3} & -6.3\times 10^{-5} & -3.2\times 10^{-2}\\
-5.2\times 10^{-9} & -5.5\times 10^{-7} & 9.9\times 10^{-1} & -1.6\\
-9.2\times 10^{-8} & -9.7\times 10^{-6} & -5.7\times 10^{-2} & 9.3\times 10^{-2}
\end{array}\right).
$
}
\end{equation}
Similar Yukawa-coupling matrix can be found for the down-type quarks.
The CKM matrix is:
\begin{equation}
\resizebox{0.47 \textwidth}{!} 
{
$ 
\lvert V_{CKM}^{331}\rvert= \left(\begin{array}{ccccc}
0.974   & 0.226  & 0.00332   & 0.000014   & 0.000048\\
0.23    & 0.97   & 0.0434    & 0.00010     & 0.000086\\
0.007   & 0.0430  & 0.997    & 0.017      & 0.0001\\
0.00073 & 0.0016 & 0.057    & 0.00099     & 6.7\times 10^{-6}\\
\end{array}\right).
$
}
\end{equation}

    The SM CKM matrix elements are produced correctly at $2\sigma$ confidence level. We have checked \cite{Huitu:2019kbm} that the neutral meson mixing bounds \cite{Bona:2007vi} are satisfied in this example. 
\\

\noindent
 {\it Conclusion.}
It is interesting that Froggatt-Nielsen -mechanism, with which the hierachical structure of fermions can be realized using gauge singlet combination of triplets as an effective flavon, can be embedded in a 331-model in which also the number of generations can be understood.
Furthermore, the scale of the flavour breaking is the same as the breaking scale of the symmetry of the model.
In order to form an effective flavon, no new scalar triplets beyond those, which are needed to generate the tree-level masses for the particles, are necessary.
We also indicate that the tree-level FCNCs are suppressed in our model.

\vspace{0.5cm}

\noindent{\it Acknowledgements.}  
KH acknowledges the H2020-MSCA-RICE-2014 grant no. 645722 (NonMinimalHiggs). NK is supported by Vilho, Yrj{\"o} and Kalle V{\"a}is{\"a}l{\"a}
Foundation.
\\

\noindent{\it Appendix.}

The down-type quark mass matrix is generated by the following terms in the Lagrangian:
\begin{eqnarray}
\mathcal{L}\supset &&\sum_{\gamma=1}^{5}(y^d_{\eta})_{1\gamma}\bar{Q}'_{L,1}\eta ~d'_{R,\gamma}
+\sum_{\alpha=2}^{3}\sum_{\gamma=1}^{5}(y^d_{\rho^{\ast}})_{\alpha\gamma}\bar{Q}'_{L,\alpha}\rho^{\ast} ~d'_{R,\gamma}\nonumber\\
&&+\sum_{\alpha=2}^{3}\sum_{\gamma=1}^{5}(y^d_{\chi^{\ast}})_{\alpha\gamma}\bar{Q}'_{L,\alpha}\chi^{\ast} ~d'_{R,\gamma}+h.c.\nonumber
\end{eqnarray}
The down-type quark mass matrix is:
\begin{eqnarray}
&&m^d_{1\gamma}=\frac{v'}{\sqrt{2}}(c^d_\eta)_{1\gamma}\epsilon^{(h^\eta)_{1\gamma}},
m^d_{\alpha\gamma}=\frac{v_1}{\sqrt{2}}(c^d_{\rho^\ast})_{\alpha\gamma}\epsilon^{(h^{\rho^\ast})_{\alpha\gamma}}\nonumber\\ 
 &&m^d_{(2+\alpha)\gamma}=\frac{v_2}{v_1}m^d_{\alpha\gamma}
+\frac{u}{\sqrt{2}}(c^d_{\chi^\ast})_{\alpha\gamma}\epsilon^{(h^{\chi^\ast})_{\alpha\gamma}},\nonumber
\end{eqnarray}
where $\alpha=2,3$, and
\begin{eqnarray}
(h^\eta)_{1\gamma}&=&q(\bar{Q}_{L,1})+q(\eta)+q(d_{R,\gamma})\nonumber\\
(h^{\rho^\ast})_{\alpha\gamma}&=&q(\bar{Q}_{L,\alpha})+q(\rho^\ast)+q(d_{R,\gamma})\nonumber\\
(h^{\chi^\ast})_{\alpha\gamma}&=&q(\bar{Q}_{L,\alpha})+q(\chi^\ast)+q(d_{R,\gamma}).\nonumber
\end{eqnarray}
 The order-one coefficients are restricted to the interval $\lvert c\rvert\in[0.5,5]$. The order-one coefficients for down-type quarks are:  
 $(c^d_\eta)_{11}= 3.1679$, 
 $(c^d_\eta)_{12}= 1.0274$, 
 $(c^d_\eta)_{13}= 1.3083$, 
 $(c^d_\eta)_{14}= 1.0222$,
 $(c^d_\eta)_{15}= -1.0835$, 
 $(c^d_{\rho^\ast})_{21}= -1.3342$, 
 $(c^d_{\rho^\ast})_{22}= 2.0875$, 
 $(c^d_{\rho^\ast})_{23}= -0.8140$, 
 $(c^d_{\rho^\ast})_{24}= 1.3463$, 
 $(c^d_{\rho^\ast})_{25}= 1.6967$, 
 $(c^d_{\rho^\ast})_{31}= -1.1293$, 
 $(c^d_{\rho^\ast})_{32}= 1.9257$, 
 $(c^d_{\rho^\ast})_{33}= 1.9138$,
 $(c^d_{\rho^\ast})_{34}=  1.8382$, 
 $(c^d_{\rho^\ast})_{35}=  -0.5945$, 
 $(c^d_{\chi^\ast})_{21}= 0.9823$, 
 $(c^d_{\chi^\ast})_{22}= -0.9842$, 
 $(c^d_{\chi^\ast})_{23}= -1.5314$, 
 $(c^d_{\chi^\ast})_{24}= 4.4024$, 
 $(c^d_{\chi^\ast})_{25}= -4.7233$, 
 $(c^d_{\chi^\ast})_{31}= 1.7299$, 
 $(c^d_{\chi^\ast})_{32}= -1.0801$, 
 $(c^d_{\chi^\ast})_{33}= -0.5451$, 
 $(c^d_{\chi^\ast})_{34}= -4.5226$ and
 $(c^d_{\chi^\ast})_{35}= -3.4529$. 
 
The order-one coefficients for up-type quarks are: 
 $(c^u_{\rho})_{11}=-3.3709$, 
 $(c^u_{\rho})_{12}=2.4799$, 
 $(c^u_{\rho})_{13}= 1.1381$, 
 $(c^u_{\rho})_{14}=1.5495$, 
 $(c^u_{\eta^\ast})_{21}=-0.6744$, 
 $(c^u_{\eta^\ast})_{22}=-1.6193$, 
 $(c^u_{\eta^\ast})_{23}=1.0648$,  
 $(c^u_{\eta^\ast})_{24}=1.3911$, 
 $(c^u_{\eta^\ast})_{31}=1.3571$,
 $(c^u_{\eta^\ast})_{32}=1.3324$, 
 $(c^u_{\eta^\ast})_{33}=0.9053$, 
 $(c^u_{\eta^\ast})_{34}=1.7678$,
 $(c^u_{\chi})_{11}=1.6150$, 
 $(c^u_{\chi})_{12}=0.6765$,  
 $(c^u_{\chi})_{13}=2.0687$ and
 $(c^u_{\chi})_{14}=1.0458$.

The scalar potential parameters used are: 
$\lambda_1=0.4$, 
$\lambda_2=0.2898$, 
$\lambda_3=0.9$, 
$\lambda_{12}=0.5$, 
$\lambda_{13}=0.5$, 
$\lambda_{23}=0.22$, 
$\widetilde{\lambda}_{12}=0.8$, 
$\widetilde{\lambda}_{13}=0.9$, 
$\widetilde{\lambda}_{23}=-0.1$ and 
$b=-(10~~\textrm{TeV})^2$.
All the scalar masses are positive with these parameters. Except for the SM-like Higgs, all the scalars are heavy with their masses being larger than 10 TeV.

\end{document}